\documentclass[prstab,showpacs,preprintnumbers,amsmath,amssymb,showkeys]{revtex4}

\usepackage{graphicx}% Include figure files
\usepackage{bm}% bold math
\begin{document}
\title{Neutral Current $\nu$ Induced Reactions in Nuclei at Supernova Neutrino Energies}
%\keywords{nuclear effects, neutrino-nucleus interactions, quasielastic scattering}
\author{S. Chauhan, M. Sajjad Athar and S. K. Singh}
\affiliation{Department of Physics, Aligarh Muslim University, Aligarh-202 002, India}
\date{\today}% It is always \today, today,
\begin{abstract}
We calculate cross sections for the neutral current induced neutrino/antineutrino reaction from $^{208}Pb$ target and applied it to study Supernova neutrino event rates.
The calculations are done in local density approximation taking into account Pauli blocking, Fermi motion effects and renormalization of weak transition strengths in the nuclear medium. 
The numerical results for the neutrino nucleus total cross sections have been averaged over the various Supernova neutrino/antineutrino fluxes available in literature.
\end{abstract}
\maketitle
\section{Introduction}
In the late phase of stellar evolution the gravitational collapse of the core of a massive star takes place in which a huge amount of energy is released over a period of few tens of seconds.
 The energy carriers are mainly neutrinos and thus called supernova neutrinos. This constitutes almost equal proportion of all the flavor of neutrinos. 
The average energy for $\nu_e$: $\langle E_{\nu_e}\rangle$ = 12 MeV, ${\bar \nu_e}$:$\langle E_{\bar \nu_e}\rangle$ = 15 MeV and 
for all other flavors $\nu_\mu$,$\nu_\tau$, $\bar{\nu_\mu}$, $\bar{\nu_\tau}$: $\langle E \rangle$= 25 MeV. 
It is expected that these neutrinos may provide valuable information about the stellar core, its equation of state and the dynamics of core collapse and supernova explosion mechanism. 
These neutrino burst from a galactic supernova can be detected in terrestrial detectors. There are presently various detectors like Super-K, MiniBooNE, IceCube which are capable of detecting supernova neutrinos. The various detectors\cite{KATE} like HALO, Icarus, LBNE LAr, LBNE WC, MEMPHYS, Hyper-K, LENA, GLACIER are planned in future which will also be sensitive to supernova neutrino detection.
 Most of these detectors are using nuclear targets like $^{16}O$, $^{40}Ar$ and $^{208}Pb$. In this paper we have calculated event rates for neutral current neutrino induced process
\begin{equation}
\nu_{l}(k) + N(p) \rightarrow \nu_{l}(k^\prime) + N(p^\prime);~~N=\rm{proton~~ or~~ neutron}
\end{equation}
for the reaction taking place in the $^{208}Pb$ target. These events are presented for 1kT of target material. When the above process takes place inside the nucleus various nuclear medium effects comes into play. We have calculated the cross section in the Fermi gas model using the local density approximation and took into account Pauli blocking, Fermi motion, renormalization of the weak transition strength in the nuclear
medium~\cite{EPJA}. The effects of Fermi motion and Pauli blocking are taking into account through the imaginary part of the Lindhard function for the particle hole 
excitations in the nuclear medium. The renormalization of the weak transition strengths are calculated in the random phase approximation(RPA) through the interaction of the p-h excitations 
as they propagate in the nuclear medium using a nucleon-nucleon potential described by pion 
and rho exchanges. 
The expression of total scattering cross section in the local density approximation inside the nucleus is given by~\cite{EPJA}
\begin{eqnarray}
\sigma(E_\nu)=-2{G_F}^2\int^{r_{max}}_{r_{min}} r^2 dr \int^{{k^\prime}_{max}}_{{k^\prime}_{min}}k^\prime dk^\prime 
\int_{Q^{2}_{min}}^{Q^{2}_{max}}dQ^{2}\frac{1}{E_{\nu_l}^{2} E_l} L_{\mu\nu}J^{\mu\nu} Im{U_N}({\bf q_{0}}, {\bf q})~~~~
\end{eqnarray}
Here $q= k - k^\prime$. $L_{\mu\nu}$ is the leptonic tensor = $\bar\sum\sum L_\mu {L_\nu}^\dagger$ where the leptonic current $L_{\mu}=\bar{u}(k^\prime)\gamma_\mu(1-\gamma_5)u(k)$. $J^{\mu\nu}$ is the hadronic tensor = $\bar\sum\sum J^\mu {J^\nu}^\dagger$ where ${J^{\mu}}$ is the hadronic current given by
\begin{eqnarray}
J^\mu = \bar{u}(p^\prime)[\left(\gamma_{\mu} - \frac{\not q q_{\mu}}{q^2}\right)\tilde F_{1}^{N} + \frac{i}{2M_{N}}\sigma_{\mu \nu}q^{\nu}\tilde F_{2}^{N} + \gamma_{\mu}\gamma_{5} \tilde F_{A}^{N} + \frac{q_{\mu}}{M_{N}}\gamma_{5} \tilde F_{P}^{N} ]u(p)
\end{eqnarray}
where $\tilde F_{1,2}^{N}$, $\tilde F_{A}^{N}$ and $\tilde F_{P}^{N}$ are the vector, axial and pseudoscalar form factors, respectively. 
These form factors are in turn defined in terms of the standard Dirac and Pauli form factors of the nucleon $F_{1}^{p,n}$ and $F_{2}^{p,n}$ and a strange component $F_{1,2}^{s}$, in the following way
\[\tilde F_{1,2}^{p} = (\frac{1}{2} - 2 sin^{2}\theta_{W})F_{1,2}^{p} - \frac{1}{2} F_{1,2}^{n} - \frac{1}{2} F_{1,2}^{s},~~\tilde F_{1,2}^{n} = (\frac{1}{2} - 2 sin^{2}\theta_{W})F_{1,2}^{n} - \frac{1}{2} F_{1,2}^{p} - \frac{1}{2} F_{1,2}^{s}\]
where $\theta_{W}$ as the weak mixing angle.

For simplicity we have taken $F_{1,2}^{s}(0)$=0 in our calculation. Axial form factors $\tilde F_{A}^{p,n}$ are given by
\begin{equation}
\tilde F_{A}^{p,n} = \pm \frac{1}{2} F_{A} - \frac{1}{2} F_{A}^{s}; F_{A}^{s}(q^2) = \frac{\Delta s}{(1 + \frac{q^2}{M_{A}^{2}})^2} \nonumber
\end{equation}
where $M_A$(=1.1GeV) is axial dipole mass, $\Delta s$ denotes the strange contribution to the nucleon spin taken as $\Delta s$ = -0.15. 
The expression for the $F_{1,2}^{p}$, $F_{1,2}^{n}$, $F_{A}$ and $F_{P}$ are taken from the Ref.\cite{EPJA}. 
The consideration of renormalization of weak transition strengths in the nuclear medium leads to modified hadronic tensor $J^{\mu\nu}_{RPA}$,
the expression for which is given in Ref.~\cite{EPJA}. 
% &&&&&&&&&&&&&&&&&&&&&&&&&& REsults &&&&&&&&&&&&&&&&&&&&&&&&&&&&&&&&&&&&&&&&&&&&&&&&&&&&&&&&&&&&&&&&&&&7
\section{Results \& Discussions}
In Fig.1, we have presented the results for the scattering cross section $\sigma$ as a function of neutrino and antineutrino energies, 
for the free case as well as in $^{208}Pb$ target obtained using the local density approximation(LDA) with and without the Random Phase Approximation(RPA) correlations. We find that the inclusion of Pauli blocking \& Fermi motion effects reduces the cross section from the free case which is around 90$\%$ at E$_\nu$ = 10-20 MeV, 82$\%$ at E$_\nu$ = 50 MeV \& 65$\%$ at E$_\nu$ = 100 MeV. The cross section further reduces by about 60-65$\%$ in the energy range 10-100 MeV when RPA effects are taken into account.
The reduction in the case of antineutrino is a bit larger. 
\begin{figure}
\includegraphics[height=.28\textheight,width=1.0\textwidth]{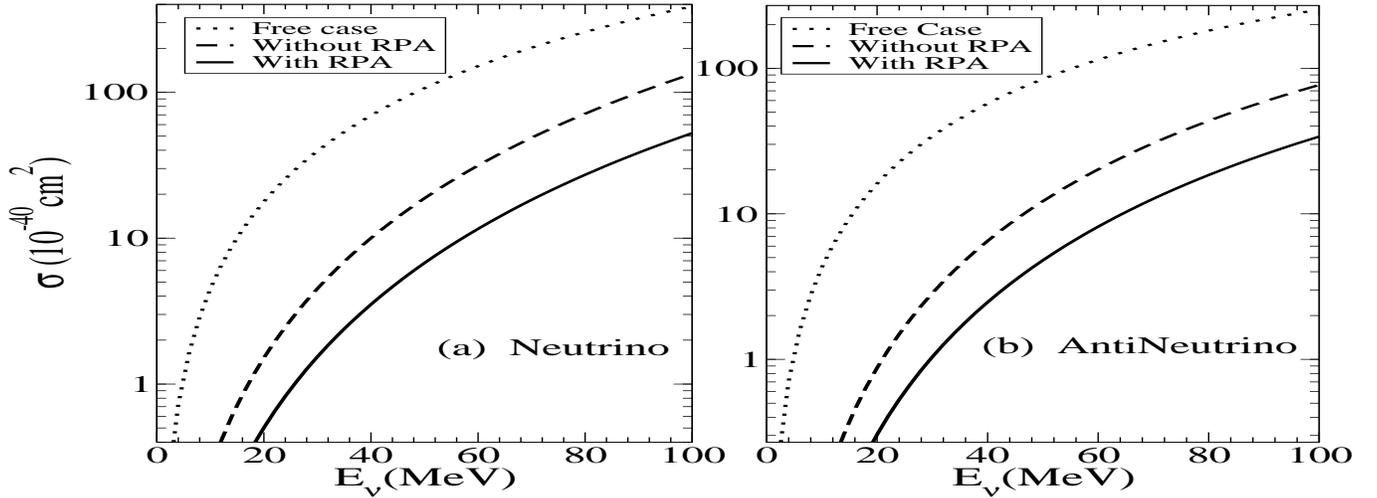}
\caption{$\nu + _Z^AX \rightarrow \nu + _Z^AX^*$ scattering cross section $\sigma(E)$ vs E for $^{208}Pb$ nucleus.}
\label{fig:nc_leadxsection}
\end{figure}
\begin{table}
%\resizebox{10.2cm}{1.8cm}{
\begin{tabular}{|c|c c|c|c|}\hline\hline
$\nu$ Flavor& ~~~~~~~~~~~~~~~~~ Duan\cite{FLUX_WEB} && Kneller\cite{Kneller}&Livermore\cite{Livermore} \\\hline
&  Without& With RPA& With RPA &With RPA\\
& RPA&&&\\\hline\hline
$\nu_e$&40 &14 &339 &315 \\
$\bar\nu_e$&119 &44 &217 &665\\\hline
$\nu_\mu$&202 &71& 337 &990\\
$\bar\nu_\mu$&41 &15 & 226& 360\\\hline
$\nu_\tau$&165 & 57 & 337 & 990\\
$\bar\nu_\tau$&96 & 35 & 217 & 665\\\hline\hline
\end{tabular}
\caption{Number of events for 1kT of lead target}
\end{table}
%&&&&&&&&&&&&&&&&&&&&&&&&&&&& Events rates &&&&&&&&&&&&&&&&&&&&&&&&&&&&&&&&&&&&&&&&&&&&&&&&&&&&&&&&&&&&&&&&&&&&&&&&&&&&&&&&&&
The event rates are calculated using the numerical value for the flux averaged cross section  
\[\langle \sigma \rangle = \int \sigma(E) \phi(E) dE,\]
where $\sigma(E)$ is the cross section shown in Fig.1, $\phi(E)$ is the supernova neutrino flux taken from Refs.\cite{FLUX_WEB,Kneller,Livermore}. 
For the 1 kT target and time taken to be 1 second, the event rates are obtained and have been shown in Table 1.  We find that with the inclusion of RPA correlations the event rate reduces by about 65$\%$. Besides the nuclear medium effects the uncertainity in event rates is also due to the flux given by various groups. 
The use of different fluxes\cite{FLUX_WEB,Kneller,Livermore} effect the event rates a lot.

\end{document}